# Evolution of CTAB/NaSal Micelles: Structural Analysis by SANS

Christopher N. Lam,*[a] William D. Hong,[a] Changwoo Do,[a] and Wei-Ren Chen*[a]

[a]Neutron Scattering Division, Oak Ridge National Laboratory, Oak Ridge, Tennessee 37831, USA.

**Abstract**: Surfactants are amphiphilic molecules that spontaneously self-assemble in aqueous solution into various ordered and disordered phases. Under certain conditions, one-dimensional structures in the form of long, flexible wormlike micelles can develop. Cetyltrimethylammonium bromide (CTAB) is one of the most widely studied surfactants, and in the presence of sodium salicylate (NaSal), wormlike micelles can form at very dilute concentrations of surfactant. We carry out a systematic study of the microscopic structures of CTAB/NaSal over a surfactant concentration range of 2.5 - 15 mM and at salt-to-surfactant molar ratios of 0.5 - 10. Using small-angle neutron scattering, we qualitatively and quantitatively characterize the equilibrium structures of CTAB/NaSal, mapping the phase behavior of CTAB/NaSal at low concentrations within the region of phase space where nascent wormlike micelles transition into long and entangled structures.

## Introduction

The amphiphilicity of surface acting agents (surfactants) lead to their self-assembly into micellar structures in aqueous solutions. These micellar aggregates exhibit structural polymorphism, governed by a multitude of factors—surfactant geometry, charge, chemistry and interactions, and external factors such as temperature, ionic strength, and flow field. While spherical micelles are generally the preferred structure for a dilute solution above the critical micelle concentration (cmc), under conditions favouring a reduction in curvature, micelles can grow into long (up to several microns)[1,2] and flexible cylindrical structures, often referred to as threadlike or wormlike micelles (WLMs).

Analogous to semiflexible polymer chains, long wormlike micelles can entangle and form a transient network, giving rise to spectacular viscoelastic properties.[3–7] In fact, Dow Chemical originally developed wormlike micelles as viscoelastic surfactants (VES).[8] Wormlike micelles have found applications in heating and cooling installations,[9] in the oil industry,[8] as drug delivery systems (using biocompatible and biodegradable surfactants),[10–12] and as viscosity modifiers in liquid detergents and personal care products. In addition to their immense practical impact, they have garnered much fundamental interest over the past few decades, as their dynamic structure renders them a good model for "living" polymers.[13–15]

Small-angle neutron scattering (SANS) has been applied in myriad structural characterization studies of micellar aggregates, including one of the most widely studied wormlike micelle systems, cetyltrimethylammonium bromide (CTAB). Approximately 30 years ago, a series of SANS investigations[16–18] of micellar solutions under shear flow, including CTAB/KBr, identified the presence of rodlike micelles. Thiyagarajan and coworkers[19] measured CTAB/NaSal at surfactant concentrations of 25 and 100 mM; although a model was unavailable at the time to incorporate the intermicellar structure factor to model the high concentration data, they determined that the system comprised polydisperse rodlike micelles at 25 mM. More recently, Das et al.[20] characterized the micellar structures of CTAB/NaSal at high surfactant concentrations (100 - 800 mM) at a fixed salt-to-surfactant molar ratio ($C_s/C_d$) of 0.6, using a combination of a cylindrical form factor with a Schulz-Zimm distribution for polydispersity in length, circular cross-sectional radius with Gaussian polydispersity, and a structure factor based on the random phase approximation (RPA) to analyse their wormlike micelle data. Building upon the scattering models for semiflexible chains developed by Pedersen and Schurtenberger,[21,22] in 2006 Chen et al.[23] reported their PRISM approach that enabled interacting micellar systems to be modelled with the micellar flexibility as a fitting parameter; using their scattering function, the effect on micellar structure due to cationic headgroup and ionic strength in CTAB and hexadecylpyridinium bromide (CPyB) micelles in the presence of NaBr were quantitatively investigated.

Knowledge of the phase behaviour of an amphiphilic system can provide insights into the thermodynamic parameters governing the properties of the system. Despite the many structural investigations of wormlike micelles using small-angle scattering, the phase behaviour of amphiphilic systems that can self-assemble into long, flexible micelles at low surfactant concentration has not been mapped in detail. CTAB/NaSal is an example of a model wormlike micelle system that is a highly effective viscosifier; the strong 1:1 complexation between CTAB and Sal$^-$ lead to the formation of long, entangled wormlike micelles at very dilute concentrations.[5] Therefore, in this study we focus on characterizing the static structure of CTAB/NaSal solutions at relatively low surfactant concentrations, where long wormlike micelles begin to form, over a wide range of salt-to-surfactant molar ratio, $C_s/C_d$. CTAB/NaSal solutions with [CTAB] = 2.5, 7.5, 10.0, and 15.0 mM, each with different ratios of $C_s/C_d$ = 0.5, 1.0, 1.75, 2.5, 5.0, and 10.0, are characterized. Using the wormlike micelle scattering function developed by Chen et al.,[23] we determine the formation and evolution of wormlike micelles both qualitatively and quantitatively, culminating in an understanding of the phase behaviour of CTAB/NaSal within this region of phase space.

## Experimental

### Materials and Samples

Hexadecyltrimethylammonium bromide (CTAB, >99.0%) and sodium salicylate (NaSal, >99.5%) were purchased from MilliporeSigma and used without further purification. Deuterium oxide (D$_2$O, >99.9% purity) was obtained from Cambridge Isotope Laboratories, Inc. Samples with surfactant concentrations ranging from 2.5 to 15 mM were prepared at salt-to-surfactant ($C_s$/$C_d$) molar ratios of 0.5, 1.0, 1.75, 2.5, 5.0, and 10.0. Samples were prepared by weighing the appropriate mass of surfactant and salt and adding the necessary volume of D$_2$O to achieve the desired concentrations. The solutions were allowed to mix for two days and equilibrate for at least one day prior to sample measurement.

### Small-Angle Neutron Scattering

SANS measurements were performed at the Extended $Q$-Range Small-Angle Neutron Scattering Diffractometer (EQ-SANS) at the Spallation Neutron Source (SNS) at Oak Ridge National Laboratory (ORNL). Two configurations were used—sample-to-detector distance of 8 m with an incident neutron wavelength of 8 Å, and 4 m and 2.5 Å —to cover a $Q$-range of 0.002 to 0.48 Å$^{-1}$. The samples were loaded into Banjo cells with a path length of 2 mm, and all the measurements were performed at 20.0 +/- 0.1 °C using a Julabo recirculating water bath. Measurements were corrected for detector background, sensitivity, and empty cell scattering, and ware normalized to absolute units using the reference scattering from a calibrated standard.

## Results and Discussion

The radially averaged $I(Q)$ for wormlike micelles is relatively featureless, displaying monotonically decaying behavior with increasing $Q$. Figure 1a shows scattering intensities at a fixed concentration of [CTAB] = 2.5 mM at different ratios of $C_s$/$C_d$; the low-$Q$ region ($Q$ < 0.006 Å$^{-1}$) shows a subtle increase in the low-$Q$ intensity with increasing salt, representing the growth of longer WLMs. Although the presence of WLMs is difficult to discern from a typical SANS intensity curve, a Holtzer or bending rod plot representation of the SANS data—which plots $QI(Q)$ vs. $Q$—clearly shows a distinct low-$Q$ peak that is a characteristic feature of long WLMs. Figure 1b shows the evolution from rods to WLMs with increasing salt in the bending rod plot representation: At a ratio $C_s$/$C_d$ = 0.5, the data are well fit by the form factor for a simple cylinder and do not exhibit a low-$Q$ peak; as salt concentration increases, the data and fit curves (to be discussed in the upcoming paragraphs) show a clear peak near $Q \sim 0.003$ Å$^{-1}$. Additionally, the bending rod plots of WLMs show a two-step decay in intensity that provides further qualitative evidence for the presence of WLMs.

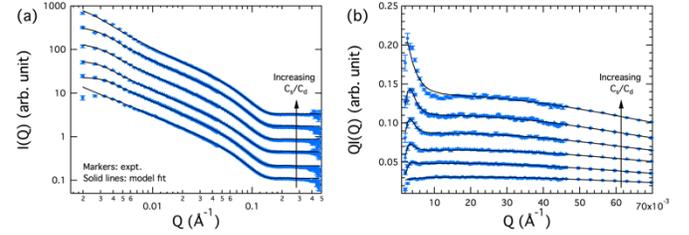

**Figure 1**. (a) Scattering curves for [CTAB] = 2.5 mM with increasing salt/surfactant ratio, $C_s$/$C_d$. The values of $C_s$/$C_d$ are 0.5, 1.0, 1.75, 2.5, 5.0, and 10.0, increasing from bottom to top. With increasing salt, the micelles evolve from stiff rods to flexible wormlike micelles, which is suggested by the slight upturn at low $Q$. (b) Same scattering curves as in (a) shown in the Holtzer, or bending rod plot, representation, $QI(Q)$ versus $Q$. In this representation, the growth of a peak feature at low Q is observed with increasing salt, demonstrating qualitatively the transition from stiff rods to wormlike micelles. Curves are offset for visual clarity.

Next, quantitative information of the local structure of WLMs can be obtained from a Guinier-like plot—$QI(Q)$ vs. $Q^2$—in the intermediate-to-high-$Q$ region, which was first proposed by Porod:[24–26]

$$QI(Q) = K\exp(-Q^2 R_{G,CS}^2) \qquad [1]$$

where $R_{G,CS}^2$ is the weight-average radius of gyration of the cylinder cross section and the value of the factor $K$ is given by

$$K = c\pi \langle N/L \rangle_w (b_m - V_m \rho_s)^2 \qquad [2]$$

In Eqn. 2, $c$ is the surfactant concentration in units of number density, $\rho_s$ is the scattering length density of the solvent, and $b_m$ and $V_m$ represent the total bound scattering length and the volume per surfactant monomer in the micelle, respectively. From Eqns. 1 and 2, the weight-average aggregation number per unit length, $\langle N/L \rangle_w$, can be calculated. Figure 2 shows the linear regression results to $\ln[QI(Q)] = \ln[K] - Q^2 R_{G,CS}^2$ in the $Q$ range corresponding to 2 x 10$^{-3}$ Å$^{-2}$ < Q$^2$ < 10$^{-2}$ Å$^{-2}$. For CTAB, $b_m$ is calculated to be -2.14 x 10$^{-4}$ Å, and $V_m$ is determined to be 345.96 Å$^3$; D$_2$O has a scattering length density of $\rho_s$ = 6.393 Å$^{-2}$. $\langle N/L \rangle_w$ is calculated to be 11.38 ± 0.65 Å$^{-1}$, showing that the weight-average aggregation number per unit length essentially remains constant as the micelles evolve with changing surfactant and salt concentration.

The magnitude of the slope from the linear regression curves in Figure 2 gives the weight-average squared radius of gyration of the cylinder cross section of the micelles. For a circular cross section, the micellar cross section radius, $R_{CS}$, is given by $R_{CS} =$ $\sqrt{2}R_{G,CS}$. The micellar cross section radius is calculated to be $R_{CS}$ = 10.80 ± 0.37 Å for [CTAB] = 2.5 mM, 14.36 ± 0.46 Å for [CTAB] = 7.5 mM, 14.84 ± 0.50 Å for [CTAB] = 10 mM, and 15.20 ± 0.59 Å for [CTAB] = 15 mM. The cross section radius should be congruent with the length of a CTAB molecule (approximately 20 Å), so these calculated values appear to be too small and show some systematic deviation from the expected value. Quantitative model fitting results yield a cross section radius of 22 Å, as will be discussed next.

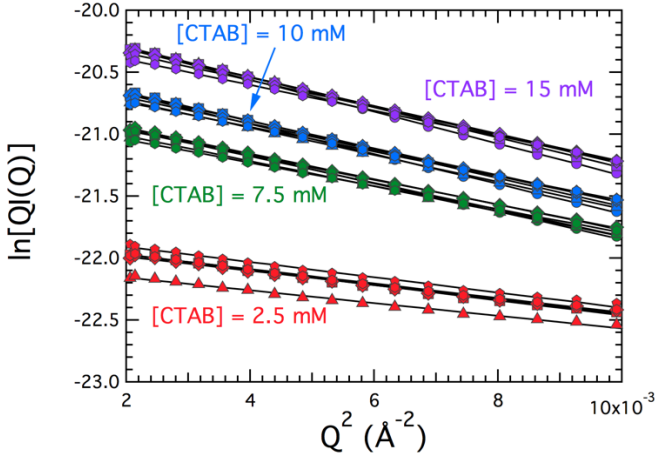

**Figure 2**. Guinier-like plot of the SANS data in the intermediate $Q$ range. The solid curves show the linear regression results to Eqn. 1, from which the intercept is used to determine the weight-average aggregation number $\langle N/L \rangle_w$, and the slope gives $R^2_{G,cs}$, the weight-average squared radius of gyration of the cross section. The high consistency of the data at different [CTAB] show that the local micellar structure—both $\langle N/L \rangle_w$ and $R^2_{G,cs}$—do not change with salt concentration. The different marker symbols represent different molar ratios of salt to surfactant, $C_s/C_d$.

Several theoretical efforts have focused on developing accurate scattering functions to model the experimental small-angle scattering data of WLMs, allowing quantitative structural information, including the contour length $L$, Kuhn length $b$, and cross section radius $R_{CS}$, to be determined. Pedersen and Schurtenberger proposed a phenomenological expression for the full scattering function of a single semiflexible chain with the incorporation of excluded volume interactions, $S_{WC}(Q,L,b)$.[21] For WLMs in which the contour length is significantly greater than the cross section radius, the scattering from the cross section can be separated from that of the contour length and Kuhn length via the decoupling approximation[27] as

$$I_{WC}(Q, L, b, R_{CS}) = c\Delta\rho_m^2 M_w S_{WC}(Q, L, b) P_{CS}(Q, R_{CS}) \quad [3]$$

where $c$ is the surfactant concentration, $\Delta\rho_m^2$ is the scattering contrast, $M_w$ is the average micelle molecular weight, and assuming a circular cross section, $P_{CS}$ is given by the scattering function for the cross section of a rigid rod:

$$P_{CS}(Q, R_{CS}) = \left[\frac{2J_1(QR_{CS})}{QR_{CS}}\right]^2 \quad [4]$$

Polydispersity of the contour length is incorporated using a Schulz-Zimm distribution with polydispersity index $z$ = 1, setting $M_w/M_n$ = 2. To model intermicellar interactions that are essential for capturing accurate micellar structural information, Pedersen and Schurtenberger proposed the following structure factor,[22] based on the polymer reference interaction site model (PRISM) and Monte Carlo simulations:

$$S_{PRISM}(Q, L, b) = \frac{S_{WC}(Q, L, b)}{1 + \beta c(Q) S_{WC}(Q, L, b)} \quad [5]$$

Here, $\beta$ is a parameter that represents the intermicellar interaction strength, and $c(Q)$ is the normalized Fourier transform of the direct correlation function for spheres on a chain and is found empirically to be well approximated by the form factor of an infinitely thin rod. Therefore, the scattering function for a wormlike micelle with intermicellar interactions is given by

$$\langle I_{WC}(Q, L, b, R_{CS})\rangle_{SZ} = c\Delta\rho_m^2 M_w \langle S_{PRISM}(Q, L, b)\rangle_{SZ} P_{CS}(Q, R_{CS}) + B_{inc} \quad [6]$$

where $\langle \cdots \rangle_{SZ}$ denotes a Schulz-Zimm distribution average and $B_{inc}$ the incoherent background.

Chen et al.[23] improved the wormlike micelle scattering model, allowing the Kuhn length to be optimized in an interacting micellar system, and this scattering function is used to fit the wormlike micelle scattering data in this study. Figure 3 summarizes the results of the structural parameters—the contour length $L$; Kuhn length $b$; ratio $b/L$, which represents the micelle flexibility; and micelle radius of gyration $R_G$—determined from model fitting of the samples that formed WLMs. The cross section radius obtained from model fitting is found to be 22.05 ± 0.10 Å, showing negligible dependence on surfactant and salt concentration. Note that this value from model fitting is larger than those calculated from the aforementioned Guinier-like plot analysis and is more consistent with the length of a CTAB molecule and with previous results.[23] With increasing CTAB concentration, it was found that a larger ratio of $C_s/C_d$ was required before the system formed long, flexible wormlike micelles. When $C_s/C_d$ = 0.5, there is insufficient salt to screen electrostatic interactions and lead to the formation of wormlike micelles, and these data points are not included in Figure 3. At the lowest CTAB concentration studied of 2.5 mM at $C_s/C_d$ = 0.5, the SANS data were well fit by a homogeneous cylinder form factor with a length of 2354.79 ± 173.11 Å and a radius of 22.94 ± 0.04 Å. At the higher CTAB concentrations of 10 and 15 mM, when $C_s/C_d$ = 0.5, 1.0, and 1.75, the SANS data were not satisfactorily fit by the wormlike micelle scattering model, and therefore, data points at these concentrations are also not shown in the plots in Figure 3. Further details of the SANS data and model fitting results are included in Supplementary Information.

In Figure 3a, we observe an overall trend of increasing micelle contour length with increasing surfactant concentration. Increasing salt also promotes the growth of wormlike micelles. NMR studies on CTAB/NaSal have revealed that the benzene ring in salicylate partially penetrates into the micelle core with the negatively charged COO$^-$ group oriented away from the micelle surface;[5,28–33] as more NaSal is added, more counterions associate with CTAB, screening the repulsive interaction between the cationic headgroup, which favors growth of the micelle length. At the lowest CTAB concentration studied, 2.5 mM, the wormlike micelle contour length increases monotonically with increasing salt, forming rather long wormlike micelles of $L \approx 5000$ Å at $C_s/C_d$ = 10.0. At a CTAB concentration of 7.5 mM, an overall trend of increasing micelle contour length with increasing salt is still observed. However, at higher surfactant concentrations, [CTAB] = 10 and 15 mM, and at relatively large ratios of $C_s/C_d$ = 2.5, 5.0, and 10.0, the contour length is determined to be approximately the same. The presence of a large excess of salt relative to surfactant decreases the micellar lifetime by increasing the rate of micellar breakage,[6,34] so it appears that when the CTAB concentration is relatively high (beyond [CTAB] = 10 mM) and at a salt-to-surfactant ratio $C_s/C_d \geq 2.5$, the equilibrium micellar contour length is limited at approximately 2700 - 3000 Å.

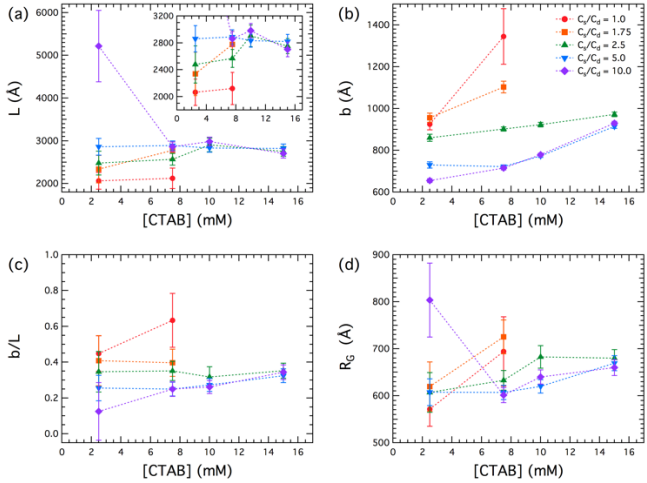

**Figure 3**. Structural parameters of wormlike micelles as a function of surfactant concentration and salt-to-surfactant ratio $C_s/C_d$: (a) contour length $L$ with zoomed-in inset, (b) Kuhn length $b$, (c) flexibility ratio $b/L$, and (d) radius of gyration $R_G$. Note that data at [CTAB] = 10 and 15 mM for $C_s/C_d$ = 1.0 and 1.75 are not shown, as the SANS data are not well fit by the wormlike micelle model due to a coexistence of rigid rods and wormlike micelles that is likely present at these conditions.

The Kuhn length shows a clear dependence on surfactant and salt concentration. At a fixed ratio of salt-to-surfactant, $C_s/C_d$, increasing surfactant concentration leads to stiffer wormlike micelles, represented by the increasing values of Kuhn length $b$ with increasing CTAB concentration in Figure 3b, similar to results observed by Chen et al.[23] Conversely, as the salt concentration increases, the wormlike micelles become more flexible, having smaller Kuhn lengths. In particular, the Kuhn length values determined for $C_s/C_d$ = 5.0 and 10.0 nearly match (at [CTAB] $\geq$ 7.5 mM), signifying that these are the inherent Kuhn length values of the wormlike micelles in the presence of a large excess of salt.

In Figure 3c, the ratio of Kuhn length to contour length, $b/L$, is summarized for different CTAB and salt concentrations. While the ratio $b/L$ shows a very weak dependence on CTAB concentration, the trend with increasing salt concentration is clear: $b/L$ decreases with increasing $C_s/C_d$, echoing the results observed for the Kuhn length in Figure 3b that the wormlike micelles become more flexible upon the addition of salt. Again the results show that the presence of excess salt beyond $C_s/C_d$ = 5.0 does not affect the micellar flexibility.

Knowing $L$ and $b$, the radius of gyration can be evaluated according to the following,[23] and the results are shown in Figure 3d:

$$\langle R_G \rangle = \sqrt{\left[\alpha\left(\frac{L}{b}\right)\right]^2 \frac{bL}{6}} \qquad [7]$$

$$\alpha(x) = \sqrt{\left[1 + \left(\frac{x}{3.12}\right)^2 + \left(\frac{x}{8.67}\right)^3\right]^{\frac{0.176}{3}}} \qquad [8]$$

When the salt is not present in large excess relative to the surfactant ($C_s/C_d$ = 1.0 and 1.75), the micelles are rather stiff (large Kuhn length), thus causing a relatively steep increase in micelle $R_G$ upon increasing CTAB concentration from 2.5 to 7.5 mM. At higher salt concentrations ($C_s/C_d \geq 2.5$), the micelle size is observed to gradually increase with increasing CTAB concentration. The $R_G$ value at [CTAB] = 2.5 mM and $C_s/C_d$ = 10.0 is calculated to be relatively large at approximately 800 Å due to the very long contour length of the wormlike micelles at this condition.

## Conclusions

CTAB/NaSal is one of the most commonly studied wormlike micellar systems that also finds use in numerous commercial applications, and by using SANS, we characterize the micellar structure within a surfactant concentration range of 2.5 - 15 mM over a wide range of salt-to-surfactant molar ratio, $C_s/C_d$ = 0.5 - 10. Results from this systematic study provide structural insight into the phase behavior of CTAB/NaSal as a function of surfactant and salt concentration (Figure 4). Results from both Guinier-like plot scaling analysis and quantitative model fitting show that the micelle cross section remains essentially independent of the surfactant and salt concentration. At all CTAB concentrations studied, wormlike micelles form, but only in the presence of sufficient salt, and they grow in length while their local structure remains the same. At very dilute CTAB concentrations and low salt concentration, rodlike micelles form that eventually grow into long and flexible wormlike micelles with increasing salt. As the surfactant concentration increases, an increasing amount of salt relative to surfactant is needed to effectively screen electrostatic interactions and allow the system to transition into wormlike micelles. In the phase region of high CTAB concentration and a relatively low amount of salt, the system is complex and consists of interacting rods, wormlike micelles, and potentially other micellar structures.

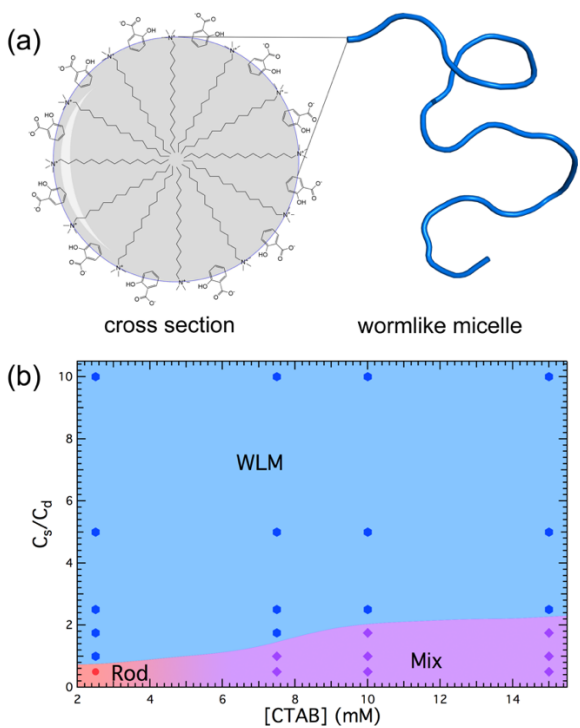

**Figure 4**. (a) Cartoon of a wormlike micelle and the cross-sectional structure, showing that on average the cross section consists of approximately 12 CTAB molecules. (b) Phase diagram of CTAB/NaSal as a function of CTAB and $C_s/C_d$. At low surfactant and salt concentration, stiff rodlike micelles form. As surfactant concentration increases, longer micelles also form, resulting in a coexistence of rods and wormlike micelles. With sufficient salt, the system is able to transition to predominantly wormlike micelles. The critical micelle concentration of CTAB in water is 0.9 mM at 25 °C.[35]

## Conflicts of interest

There are no conflicts to declare.

## Acknowledgements

This research at SNS of Oak Ridge National Laboratory was sponsored by the Scientific User Facilities Division, Office of Basic Energy Sciences, U.S. Department of Energy. W.D.H. acknowledges the support from the ORNL HSRE program.